\documentclass[conference]{IEEEtran}

\usepackage{multicol}

\usepackage{color}
\usepackage{xcolor}
\usepackage{soul}
\usepackage{amsmath}
\usepackage{graphics}
\usepackage{setspace}
\usepackage{cite}
\usepackage{latexsym}
\usepackage{float}
\usepackage{epsfig}
\usepackage{multirow}
\usepackage{cite,cases,url}
\usepackage{amssymb}
\usepackage{graphicx}
\usepackage{epstopdf}
\usepackage{balance}
\usepackage{subcaption}
\usepackage{enumerate}
\usepackage{array}
\usepackage{enumitem}
\usepackage{dblfloatfix}
\usepackage[normalem]{ulem}
\useunder{\uline}{\ul}{}
\usepackage{amsmath}
\usepackage{algorithm}
\usepackage{enumitem}
\usepackage{caption}
\usepackage{subcaption}
\usepackage{tikz}
\usepackage{fancyhdr}
\usepackage{adjustbox}
\usepackage{etoolbox}
\usepackage[noend]{algpseudocode}
\makeatletter
\usepackage{tikzpagenodes}      
\usepackage[normalem]{ulem}
\usepackage[linesnumbered,ruled,vlined,algo2e]{algorithm2e}
\useunder{\uline}{\ul}{}

\newcolumntype{L}{>{\centering\arraybackslash}m{5cm}}
\newcolumntype{K}{>{\centering\arraybackslash}m{6cm}}
\newcolumntype{P}{>{\centering\arraybackslash}m{2.3cm}}
\newcolumntype{M}{>{\raggedright\arraybackslash}m{2cm}}
\newcolumntype{N}{>{\raggedright\arraybackslash}m{2.5cm}}

\usepackage{changepage}
\fancypagestyle{firstpage}
{
    \fancyhead[L]{\footnotesize This article has been accepted for publication in the IEEE Wireless Communications and Networking Conference. Please cite it as A. S. Abdalla, A. Behfarnia, and V. Marojevic, ''Aerial Base Station Positioning and Power Control for Securing Communications: A Deep Q-Network Approach,'' IEEE Wireless Communications and Networking Conference, April 2022. } 
    \fancyhead[R]{\footnotesize \thepage}
    \cfoot{}
    
}

\begin{document}


\title{{Aerial Base Station  Positioning and 
Power Control 
for Securing 
Communications: A Deep Q-Network Approach}}  

\author{
\IEEEauthorblockN{Aly Sabri Abdalla\IEEEauthorrefmark{1}, Ali Behfarnia\IEEEauthorrefmark{2}, and
Vuk Marojevic\IEEEauthorrefmark{1}
}\\ \vspace{-0.5 cm}
\normalsize\IEEEauthorblockA{\IEEEauthorrefmark{1}Department of Electrical and Computer Engineering, Mississippi State University, MS 39762, USA\\}
\normalsize\IEEEauthorblockA{\IEEEauthorrefmark{2}Department of Engineering, University of Tennessee at Martin, TN, USA\\}

Email: asa298@msstate.edu, a.behfarnia@tennessee.edu, vuk.marojevic@msstate.edu 
}
\vspace{-2 cm}
\maketitle
\thispagestyle{firstpage}

\begin{abstract}
The unmanned aerial vehicle (UAV) is   
one of the technological breakthroughs that supports a variety of services, including communications. 
UAV will play a critical role in enhancing the physical layer security of wireless networks. This paper defines the problem of eavesdropping on 
the link between the ground user and the UAV, which serves as an aerial base station (ABS). The reinforcement learning 
algorithms Q-learning and deep Q-network (DQN) are proposed for optimizing the position of the ABS and the transmission power to enhance the data rate of the ground user. 
This increases the secrecy capacity without the system 
knowing the location 
of the eavesdropper. 
Simulation results show fast convergence and the highest secrecy capacity of the proposed DQN compared to Q-learning and baseline approaches.

\end{abstract}

\IEEEpeerreviewmaketitle

\section{Introduction}
\label{sec:intro}

Unmanned aerial vehicles (UAVs) support 
various applications
in advanced 
cellular networks. 
A 
UAV can be 
an aerial base station (ABS), aerial relay (AR), or aerial user equipment (AUE) in the cellular network. As a network support node, it can enhance the performance of the end users. 
Steps to identify the challenges and solutions of emerging cellular networks serving UAVs are being taken by the 3rd Generation Partnership Project (3GPP)~\cite{3GPP_Stnds}. Major challenges are radio frequency (RF) interference and security. 
Security is increasingly important in modern wireless networks and 
needs to be ensured for establishing communications links between UAVs and terrestrial nodes~\cite{Sec, coverage, Spect}.  


Different types of attacks have been studied with the aim of strengthening
the security of wireless communications links with trusted cellular network-connected UAVs. 
Eavesdropping is a passive attack that can compromise the 
confidentiality and privacy of 
control and user data. 
Early research \cite{hovering, moving} considering both hovering and moving UAVs 
study the performance of an AR for safeguarding the ground communications links between terrestrial nodes against eavesdropping attacks.
Reference~\cite{TVT} 
investigates the same problem but for multiple ground users and multiple eavesdroppers by optimizing the position of the AR and the transmission power. In~\cite{ABSPositiononly}, the UAV is deployed as an ABS to serve ground users under 
attack with the goal of optimizing the 3D position of the UAV for maximizing the secrecy rate. 
The authors of~\cite{ABSPositiononltwihoutEVSlocation} maximize the secrecy rate of legitimate users while the ABS position is optimized without eavesdropper location information. 

In other lines of work, learning approaches have been used for solving the above problem. Reference~\cite{RLAR} develops model-free reinforcement learning (RL) algorithms to maximize the system secrecy rate by transmitting artificial noise with optimized beamforming from the AR. A multi-agent
deep RL has been proposed in~\cite{MADRL} to maximize the secrecy capacity by jointly optimizing the trajectory, transmit power, and jamming power for both relay and friendly jamming UAVs protecting against eavesdroppers with known locations. Nevertheless, limited study items utilized RL in secrecy rate analysis, such as ~\cite{RLAR,MADRL}. These approaches assume knowledge of the eavesdroppers and their locations. Also, learning methods are only recently being explored for physical layer security. 
To the best of our knowledge, none of recent studies have explored the potential of RL solutions without assuming the availability of imperfect or perfect location information of the eavesdropper. 

In this paper, we propose a RL algorithm for an ABS 
to assist with the uplink transmission of ground users that are subject to 
eavesdropping. Our solution is based on a 
{deep Q-network (DQN)} that aims to maximize the secrecy capacity by optimizing the legitimate capacity of the ground user. This is achieved by finding the position of the ABS and the transmission power that maximizes the data and secrecy rates without the eavesdropper's location information.  

The rest of the paper is organized as follows. Section II provides the system model and problem formulation. Section III introduces the  DQN as our learning-based physical layer security solution to eavesdropping. Section IV presents the numerical analysis and Section V derives the conclusions.


\section{System Model and Problem Formulation} 
\begin{figure}[t]
 \vspace{-1mm}
     \centering
     \includegraphics[width=0.40\textwidth]{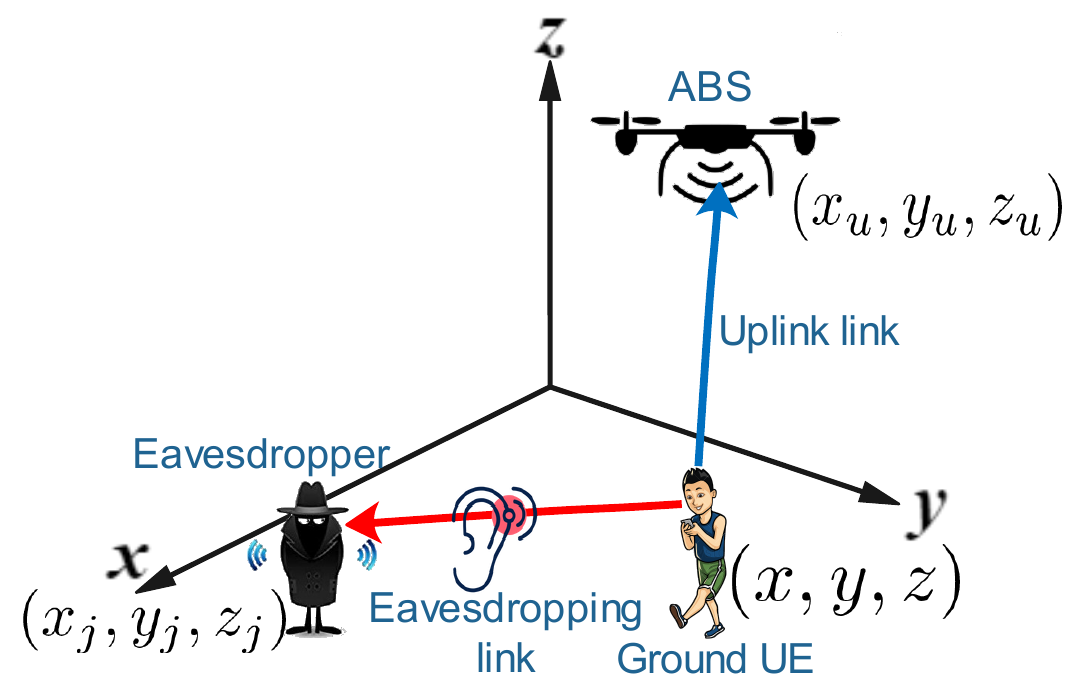}
     \vspace{-1mm}
     \caption{The simulation scenario.}
     \vspace{-6mm}
         \label{fig:system}
 \end{figure}

\vspace{-1 mm}
\label{sec:system}
The scenario 
studied in this paper is illustrated in~Fig.~\ref{fig:system}. The ground user communicates with the ABS 
in the presence of a terrestrial eavesdropper.  
The user and eavesdropper are independent of each other. 
The ABS is positioned to provide a secure and reliable uplink (UL) 
for the user under attack. It achieves this by leveraging its 3D mobility and strong line of sight (LoS) channel that allows low power transmission. 
Without loss of generality, we model and analyze the UL channel here. The same principles can be applied for the downlink (DL).
In the rest of this paper, we use the UAV and ABS interchangeably. 

We define the location of the ABS, the eavesdropper, and the user in the 3D Cartesian coordinate system as ($x_u,y_u,z_u$), ($x_j,y_j,z_j$), and ($x,y,z$), respectively.
For practicality, 
time slots are used to capture the different radio frames and statistical channel conditions, as well as the momentarily static position of the nodes. 
Therefore, the ABS coordinates in time slot $t$ are expressed as ($x_u[t], y_u[t], z_u[t]$).    

\subsection{Communications Channel} 
The received signal at the ABS when the ground user equipment (UE) 
transmits the signal $s$ with power $P$ is as follows
\begin{equation}
    {r} = \sqrt{P\hbar}s+n,
\end{equation}
where $\hbar$ is 
the air-to-ground (A2G) channel 
gain between the UE and the ABS, and $n$ denotes the additive white Gaussian noise (AWGN) of zero mean and $\sigma^2$ variance. %
Based on the measurements presented in~\cite{R4}, 
the LoS 
model is a good approximation for the A2G channel in rural areas. 
The channel gain 
has 
a path loss exponent of two and can be written as
\begin{equation}
\small
\begin{aligned}
    \hbar[t] &=\zeta_{0}d^{-2}_{u}[t] \\
    &=\frac{\zeta_0}{(x[t]-x_u[t])^2+(y[t]-y_u[t])^2+(z[t]-z_u[t])^2},
    \end{aligned}
    \label{eq:channel_gain1}
\end{equation}
where $\zeta_{0}$ is the channel gain at the reference point $d_0$ = 1 m. Parameter $\zeta_{0}$ is the same for the A2G and the ground-to-ground (G2G) channel 
because of the same the antenna gains and carrier frequency in both cases. Parameter $d_{u}[t]$ denotes the distance between the ground user and the ABS. 

The capacity of the channel between the UE and the ABS in time slot $t$ is calculated as follows: 
\begin{equation}
    {C _{u}[t]} = \log_2 \left( 1+\frac{{P}[t]\hbar[t]}{\sigma^2}\right) = \log_2 \left( 1+\frac{\zeta_0{P}[t]}{d^2_{u}[t]\sigma^2}\right),
\label{eq:cap_UAV-UE}
\end{equation}
where $\zeta_0/\sigma^2$ is the signal-to-noise ratio (SNR) at the reference point. 

\subsection{Eavesdropping Channel} 
Often, UEs have fewer antennas 
and transmit in broadcast mode without beamforming. 
The eavesdropper is intercepting the transmitted signal from the ground UE and the received signal can be formulated as
\begin{equation}
    {v } = \sqrt{P{\emptyset}}s+w,
\end{equation}
where $w$ denotes the zero-mean 
AWGN with a variance of $\sigma^2$ 
at the eavesdropper. 
As a result of the G2G communications link between the UE and the eavesdroppers, the exponent of the path loss for this link is assumed to be four. Within the time slot $t$, the G2G channel 
gain represented by $\emptyset$ between the UE and eavesdropper can be modeled as
\vspace{-0.5 mm}
\begin{equation}
\small
\begin{aligned}
    \emptyset &=\zeta_{0}d^{-4}_{j}[t]\\
    =&\frac{\zeta_0}{({(x[t]-x_j[t])^2+(y[t]-y_j[t])^2+(z[t]-z_j[t])^2)}^2},
    \end{aligned}
\end{equation}
where $d_{j}[t]$ corresponds to the distance between the 
UE and the eavesdropper in the $t$th time slot.
The corresponding capacity of the wiretap channel is 
\begin{equation}
    {C _{j}[t]} = \log_2 \left( 1+\frac{{P}[t]{\emptyset}[t]}{\sigma^2}\right) = \log_2 \left(
    1+\frac{\zeta_{0}{P}[t]}{d^{4}_{j}[t]\sigma^2}\right).
\end{equation}
\subsection{Secrecy Capacity Metric}
The security of the system is evaluated using the secrecy capacity metric which is 
commonly employed in the literature for analyzing 
eavesdropping security problems. 
The secrecy capacity is the rate at which the malicious node cannot decode any 
data 
when the legitimate channel capacity is higher than the wiretap channel capacity~\cite{YanWanGer2015}. 
The secrecy capacity over $T$ 
time slots 
is then obtained as
\begin{equation}
    {C_{sec}} = \frac{1}{T}\sum\limits_{t=1}^T \bigg(
    C_{u}[t]- C_{j}[t] \bigg)^+,
    \label{equ:totalsec}
\end{equation}
where $[\omega]^+ \triangleq max(\omega,0)$.

\subsection{Problem Formulation }
The optimization problem 
is defined 
according to the following 
assumptions: 
First, for the sake of simplicity and without loss of generality, we assume that an ABS flies at a constant height that 
facilitates a LoS link between the ABS and the ground UE. In general, the lower the UAV height can be, the higher the resulting legitimate channel and secrecy capacity in the considered context. Second, we assume that the location of the passive eavesdropper is unknown. 
This scenario is of interest in practice where it is difficult to detect the presence and location of eavesdroppers because of their passive nature. 
Thus, we consider the capacity of the legitimate user in the above problem formulation. 
The eavesdropper location is used only for the calculation of the resulting secrecy rate to evaluate the performance of the proposed solution. 

The objective of this paper is thus to maximize the legitimate capacity $C_{u}$  
by selecting the 
position for the ABS and controlling transmit power 
for the UE.   

The optimization of the UE capacity will result in an improved secrecy capacity of the system. 
However, the UL capacity of the ground user relies on having a short distance to the ABS  
with a strong LoS link. 
Therefore, the ABS position constraint is formulated as follows:
\begin{equation}
    (x_u[t],y_u[t]) \leq (L_x,L_y), \forall {t},
\end{equation}
where $(L_x,L_y)$ represent the maximum 2D coordinates of the UAV ground location projection. 
We introduce the peak transmission power $ P_{max}$ and the UL transmission power constraint for the legitimate UE per time slot as  
\begin{equation}
   0 \leq P[t] \leq P_{max}, \forall {t}.
   \label{11}
\end{equation}
The optimization problem of the UE capacity 
is then given as
\begin{equation}
\begin{aligned}
   (P1) : \max\limits_{x_u,y_u,P} \frac{1}{T}\sum\limits_{t=1}^T \Bigg[ 
     &\log_2 \left( 1+\frac{\zeta_0{P}[t]}{d^2_{u}[t]\sigma^2}\right)
   \Bigg]
   ,\\
   s.t.\hspace{2 pt }(8),(9).
   \end{aligned}
   \label{eq:opt_secrecy_rate}
\end{equation}
where $x_u$ and $y_u$ are the UAV positioning parameters and $P$ is the uplink transmission power controlled by the ABS. 

\section{Proposed Solution} 
\vspace{-1 mm}
\label{sec:solution}

The optimization problem (\ref{eq:opt_secrecy_rate}) is challenging because it needs a joint UAV positioning and UE transmission power adjustment in the presence of an eavesdropper. 
Since the objective function 
is non-convex with respect to parameters $x_u$, $y_u$, and $P$, and the 
constraints, the problem becomes NP-hard \cite{TVT} \cite{MADRL} \cite{li2018deep}. Alternatively, the ABS position and the UE transmission power will be selected 
through a transition process based on the current system state. Since the next state of the system is independent from the previous state and action, the process can be modeled as a Markov decision process (MDP). This allows applying a RL algorithm to a UAV agent without requiring the knowledge of the system model. In this regard, instead of solving the problem using conventional optimization algorithms
, we apply a RL method that can solve the problem in an efficient and accurate way, and thereby improve the 
secrecy rate in the network.


In what follows, we first describe the MDP model by defining the settings that include states, actions, and the reward for the UAV agent. Then, we introduce the Q-learning method based on the defined MDP settings to solve the problem. In order to avoid intractably high dimensionality for the high state-action space, we 
propose the 
DQN method in which a deep neural network (DNN) is employed to estimate the action value function for the UAV agent. 

\subsection{MDP Settings}
The MDP for the UAV agent is composed of the state space $\mathcal{S}$, the action space $\mathcal{A}$, the reward space $\mathcal{R}$, and the transition probability space $\mathcal{T}$, i.e., $\mathcal{(S, A, R, T)}$. At time slot $t$, the agent observes the state $s_t \in \mathcal{S}$, and based on its policy, it takes an action $a_t \in \mathcal{A}$. Depending on the distribution of the transition probability $\mathcal{T}(s_{t+1}|s_t, a_t)$, the agent will be transferred to the new state $s_{t+1}$. Since the transition probability is highly dependent on a specific environment and is difficult to obtain, we choose the Q-learning method as a model-free algorithm to directly find the best policy for each action in each state. This means that we do not need to know $\mathcal{T}$, but we need to carefully define states, actions, and the reward of the agent as follows.

\textbf{State:} The set of states is defined as 
\begin{equation}
\mathcal{S} = \{s_1, s_2, ..., s_t, .., s_T\},
\label{equ:MDP_state_set}
\end{equation}
where $T$ is the total number of time slots. Each state $s_t$ at a time slot $t$ has three elements that are defined as
\begin{equation}
s_t =\{\Delta x, \Delta y, \Delta z \}, 
\label{equ:MDP_states}
\end{equation}
where $\Delta x, \Delta y$, and $\Delta z$ represent the distance difference between the UAV and UE along the $x, y$, and $z$ axes, respectively. 
It is worth noting that the value of each state affects the channel gain and, hence, the SNR.

\textbf{Action:} The states are transited according to the defined actions. A set of actions is defined as
\begin{equation}
\mathcal{A} = \{a_1, a_2, ..., a_t, .., a_T\},
\label{equ:MDP_actn_set}
\end{equation}
where each action 
at time $t$ 
consists of two parts related to the UAV movement and one part related to the transmit power adjustment. That is, 
\begin{equation}
a_t = \{\delta_x, \delta_y, \delta_p\},
\label{equ:MDP_Actions}
\end{equation}
where $\delta_x$ and $\delta_y$ represent the movement in the $x$ 
and $y$ directions and $\delta_p$ denotes the change in power. 
The altitude of the UAV is assumed to be constant. 

As a sample configuration, the movement in $x$ and $y$, i.e., $\delta_x$ and $\delta_y$, can be assumed to change by $+1$ unit or $-1$ unit, and the power level, i.e., $\delta_p$, can be assumed to change by $p_1$, $0$, or $-p_1$, where $p_1$ is an arbitrary number. Hence, here we consider $4$ possible directional movements and $3$ power level changes, resulting in $12$ possible actions for the ABS, which controls the UE transmission power, in any state. 


 \textbf{Reward:} After taking an action $a_t$ in a state $s_t$ at time slot $t$, the UAV agent will receive a reward $R_t(s_t, a_t)$. The UAV should get more rewards for the actions that may lead it higher secrecy rates. In this respect, we define the reward function of the system based on the instantaneous SNR between the UAV and the UE as
\begin{equation}
    {R_t(s_t, a_t)} =  \frac{\zeta_0{P}[t]}{d^2_{u}[t]\sigma^2}. 
    \label{eq:reward}
\end{equation} 


\subsection{Q-Learning Method}
The UAV agent can apply the Q-learning method to find the best policy for the state-action relationship. Q-learning is a classical table-based RL algorithm in which the state-action pair has the value of $Q(s,a)$. The rows and columns of the Q-table consist of the environmental states and 
possible actions of the UAV agent, respectively. For example, for the aforementioned sample of states and actions, the table has $3$ rows and $12$ columns. The Q-values in the table are initially filled by random numbers. Then, the Bellman equation is used to obtain the optimal state-action pairs in the table \cite{Survey17deep}:
\begin{flalign}
Q^{*}(s, a) = E_{s^\prime}\bigg[R(s, a) + \gamma \times \max_{a\in \mathcal{A}} \ Q(s^\prime, a^\prime)\bigg],
 \label{eq:Bellman}
 \end{flalign}
where the $s^\prime$ and $a^\prime$ symbolize the next state and action. The parameter $\gamma \in (0,1)$ denotes the discount factor that affects the importance of the future reward. 
The Bellman equation is used through an iterative process to update the Q-values, where
a learning rate parameter $\alpha$ is applied to determine how quickly an agent leaves the previous Q-value for the new Q-value in the table. That is, $Q^{\text{new}}(s, a) = (1-\alpha) \ Q_{i-1}(s, a) + \alpha \ Q_{i}(s, a)$, where subscript $i$ indicates the $i$-th iteration in the process. Finally, in each step, the Q-value is updated while applying the $\epsilon-$greedy algorithm. This algorithm is used to balance between the exploration and the exploitation of 
the environment \cite{Survey17deep}.


Although Q-learning provides a general framework for RL, it requires to store the Q-values for each state-action pair in the table. The number of state-action pairs in the table grows quickly with the number of states and actions. 
As the Q-table becomes large
, the 
process 
becomes more time-consuming and 
impractical. 
Therefore, we consider the DQN method where a DNN is used to estimate the $Q(s, a)$ values
, as opposed to Q-learning, where the Q-table is used to estimate the state-action values. This has advantages in terms of handling a large number of states and actions and the associated processing time.


%
%
\begin{figure}[t]
  \centering
  \includegraphics[ height=6.2 cm]{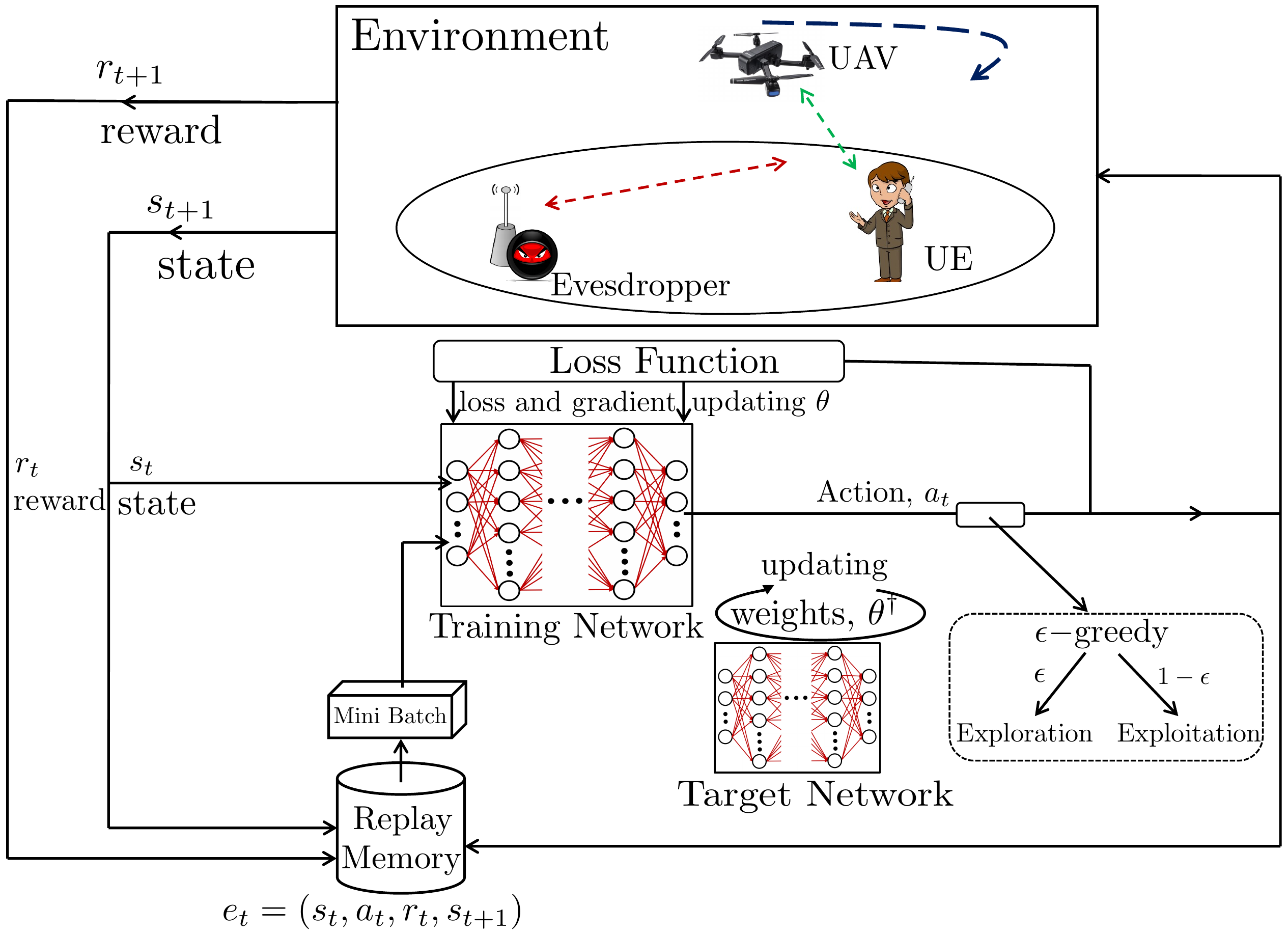}
  \caption{Block diagram of the proposed DQN architecture.}
  \label{fig:DQL}
  \vspace{-6mm}
\end{figure}
\subsection{Deep Q-Network Method}
The DQN
, initially proposed by Google Deep Mind \cite{DeepMind2015}, integrates the RL and deep learning methods. This technique uses the power of nonlinear functions, specifically DNNs, in order to approximate the Q-values and handle highly dimensional state-action problems. 

Figure \ref{fig:DQL} shows the block diagram of the proposed DQN method. 
There are two DNNs of the same structure: a training network and a target network. The training network outputs the Q-values associated with the actions of the UAV in each state. The target network supervises the training network by providing the target Q-values obtained from the Bellman equation. The target values are compared with the outputs of the training network to minimize the loss function described below. Also, the target network prevents the learning model from suffering from the noise in the environment. The inputs to the networks are the states of the UAV agent in the environment, see (\ref{equ:MDP_state_set})-(\ref{equ:MDP_states}). 
The outputs of the network are the Q-values corresponding to the actions of the UAV agent, i.e., $Q(s, a; \theta)$, where $\theta$ denotes the weights of the DNNs.

As the UAV takes an action
, the system generates a record of experience. At time step $t$, the experience contains the current state $s_t$, the action $a_t$, the reward $r_t$, and the next state $s_{t+1}$, formed as a tuple $e_t = (s_t, a_t, r_t, s_{t+1})$. Each such experience is stored in a replay memory with the capacity of $N$, such that $\mathcal{M}=\{e_1, ..., e_t, ..., e_N \}$. The memory is a queue-like buffer that stores the latest N experience vectors. 
We use a mini-batch sample from the replay memory to feed the input of the training network as shown in Fig.~\ref{fig:DQL}. The main reason for using the mini-batch samples from the reply memory is to break possible correlations between sequential states of the environment, and thereby facilitate generalization.

In order to minimize the error prediction of the DNNs, a loss function is used that is defined as 
\begin{flalign}
\notag L(\theta) = \mathbb{E} \Bigg[ \bigg( \Big[ r_t + \gamma \times \max_{a\in \mathcal{A}} \ Q(s_{t+1}, a_{t+1}; \theta^{\dagger})\Big] - \\ 
\Big[Q(s_{t}, a_{t}; \theta) \Big] \bigg)^2 \Bigg],
\label{eq:loss}
\end{flalign}
where the Q-value of the first term is obtained from the target network and the Q-value of the second term 
from the training network. 
Parameters $\theta^{\dagger}$ and $\theta$ denote the weights of the target network and training network, respectively. The $\theta^{\dagger}$ coefficients are updated every few time slots in order to ensure the stability of the target values and, hence, facilitate stable learning. 

The UAV applies a gradient descent algorithm,
\begin{flalign}
\notag \nabla_{\theta}\, L(\theta) &= - \mathbb{E} \Bigg[ 2 \ \nabla_{\theta}Q(s_{t}, a_{t}; \theta) \bigg( \, r_t + \, \gamma \ \times \\ & \qquad \max_{a\in \mathcal{A}} \ Q(s_{t+1}, a_{t+1}; \theta^{\dagger})  - Q(s_{t}, a_{t}; \theta) \bigg)  \Bigg],
\label{eq:GDS}
\end{flalign}
to update $\theta$ an
$\theta^{\dagger}$ as the weights of the DNNs with the aim of minimizing the prediction error.

Finally, we apply the $\epsilon-$greedy algorithm to select an action while balancing the exploration and the exploitation of the UAV in the environment (Fig.~\ref{fig:DQL}). In this algorithm, the UAV explores the environment with the probability of $\epsilon$ by choosing a random action. 
More precisely, the UAV exploits the environment with the probability of $1-\epsilon$ by choosing the actions that maximize the Q-value function, i.e., $a^{*} = \text{argmax}_{a \in \mathcal{A}} \ Q(s,a; \theta)$. A high value of $\epsilon$ is initially set in the model for the UAV to spend more time for the exploration. As the agent obtains more knowledge about the environment, the $\epsilon$ value is gradually decreased to leverage the experience and choose the best actions for the UAV, rather than continuing with the exploration. 

The details of the DQN-based algorithm used by the UAV agent for optimizing the UE SNR and calculating the secrecy rates is presented in Algorithm 1. The brief description of the pseudocode is as follows: The parameters of the algorithm are initialized in lines $1$ to $4$. Line $5$ starts the first loop for $K$ episodes. The environment is reset in line $6$ to initialize the starting state. The second loop begins at line $7$, representing $T$ time slots for the UAV to adjust its trajectory and power. Line $8$ denotes the state in each time slot, and lines $9$ to $13$ apply the $\epsilon-$greedy algorithm to balance the 
exploration versus exploitation. The UAV takes an action in line $14$. Then it receives the reward and goes to the new state, as denoted in line $15$. The replay memory collects the new experience in line $16$. Using the mini-batch in line $17$, the training DNN is trained in line $18$. The weights of the training DNN are updated in line $19$ using the gradient descent algorithm on the loss function of (\ref{eq:GDS}). Line $20$ updates the weights of the target DNN every $B$ time slots. Once the algorithm runs out of time slots in each episode, it updates the value of $\epsilon$ (line $21$). The rewards for each episode are stored for each episode according to line $22$.


 \begin{figure}[H]
     \centering
     \includegraphics[width=0.48\textwidth]{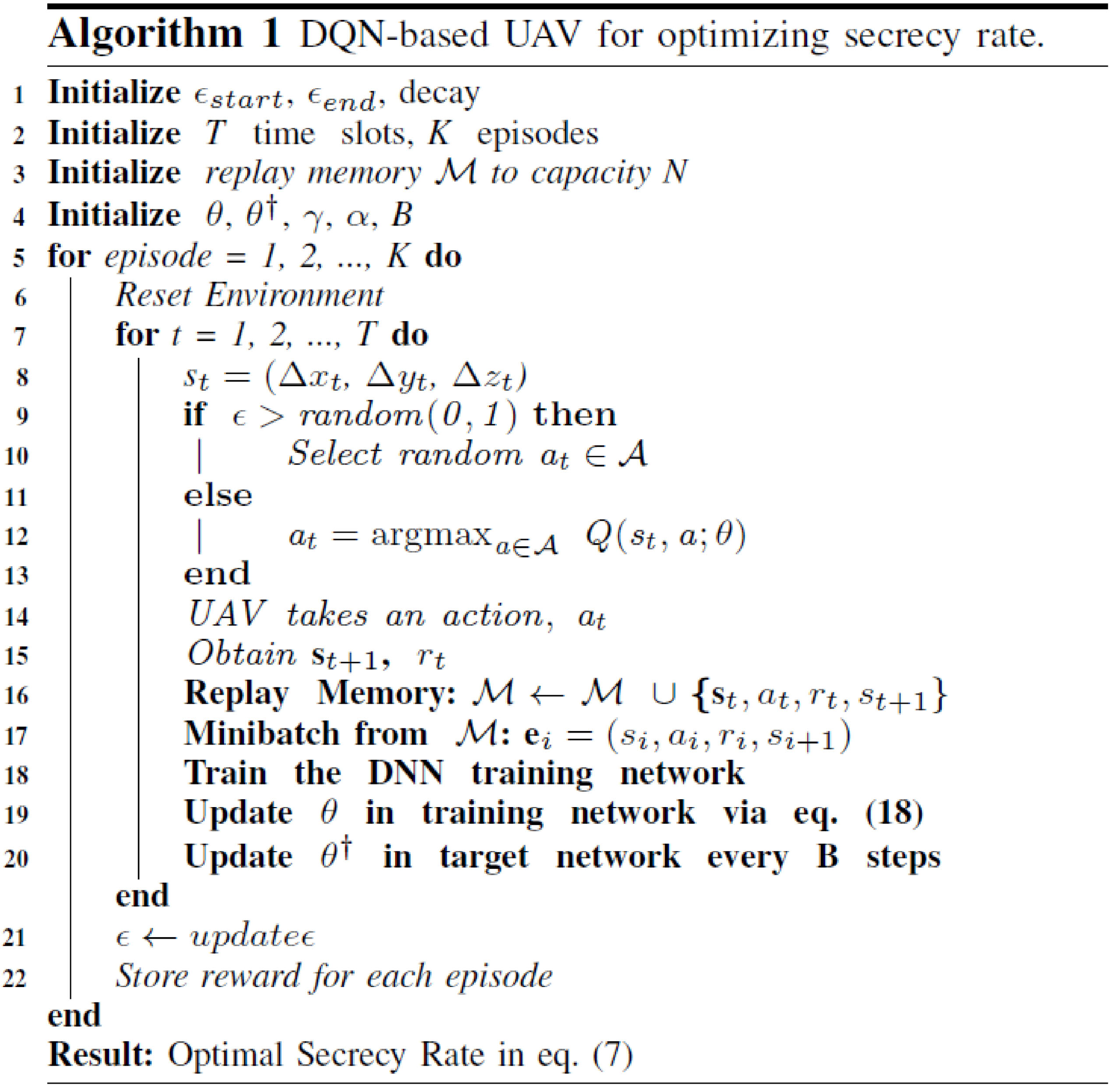}
     \vspace{-5mm}
     \label{fig:Table1}
 \end{figure}

\color{black}
\color{black}

\section{Numerical Analysis and Discussion}
\vspace{-1 mm}
\label{sec:results}
We numerically analyze the performance of the proposed DQN-based UAV positioning and power control scheme in optimizing the UL SNR of the ground user and its effect on the secrecy capacity in the presence of an eavesdropping attack. 

The simulation scenario consists of a single antenna ground UE, an ABS mounted on the UAV, and a single antenna malicious node that is performing a passive eavesdropping attack on the UL transmission 
(Fig.~\ref{fig:system}). 
The UE and the eavesdropper are randomly distributed in a 2D area that has a $L_x$ length and a $L_y$ width. The ABS is launched at a random location with a fixed altitude and is equipped with an omnidirectional antenna to enable communications with 
UEs.

Table I provides the simulation parameters and the hyperparameters for the proposed DQN solution. 
The training and target DNN networks consist of $4$ 
layers where each DNN contains two fully connected {hidden} layers, one with $24$ and the other with $32$ neurons. 
Each DNN has $3$ neurons at the input layer and $12$ neurons at the output layer, corresponding to the number of states 
and possible actions 
defined in~(\ref{equ:MDP_states}) and (\ref{equ:MDP_Actions}), respectively. 
The simulator 
is implemented in Python, using PyTorch to train the DQN. 

For the performance evaluation of the proposed DQN algorithm, we compare the resulting secrecy capacity of the DQN against the values resulting from employing Q-learning, a greedy policy, and a random state selection scheme. 
The random scheme selects the next position and power action at each time slot randomly, whereas the greedy policy assigns the next action based on the highest Q-value. 

 \begin{figure}[t]
     \centering
     \includegraphics[width=0.48\textwidth]{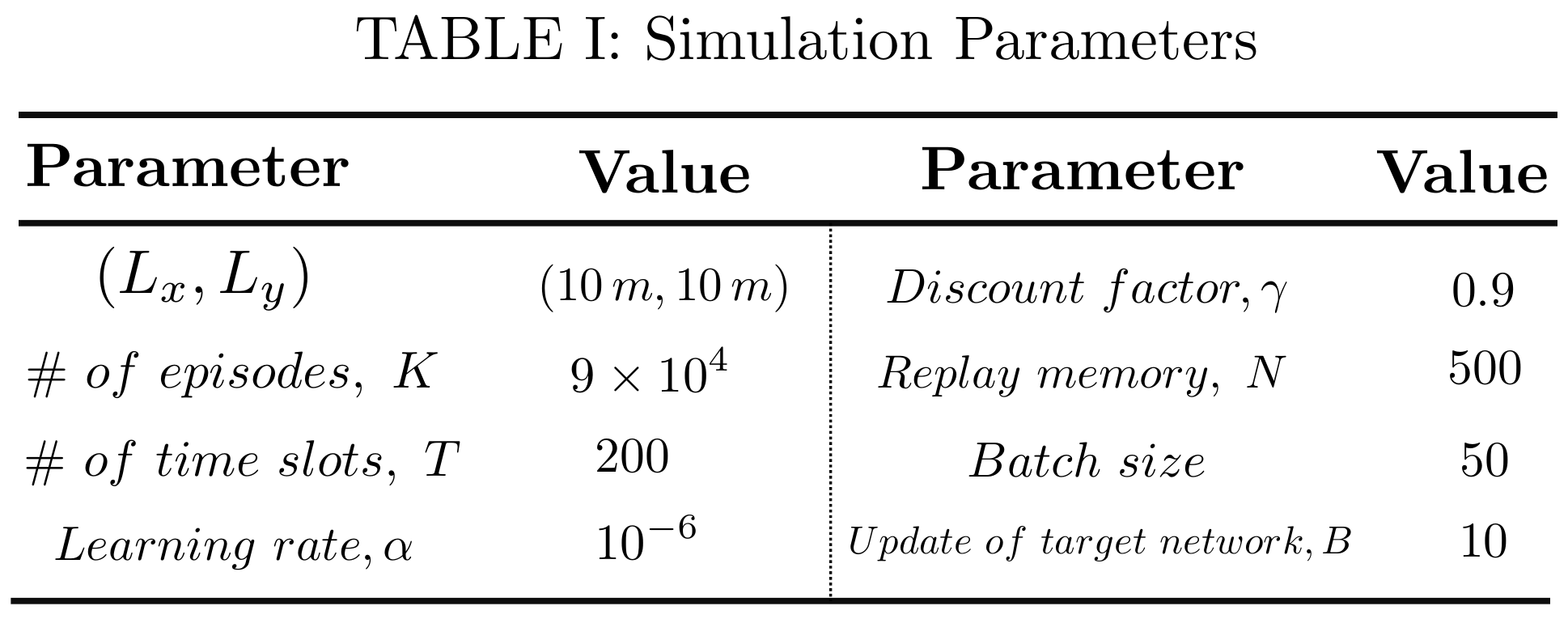}
     \vspace{-5mm}
     \label{fig:Table}
 \end{figure}

    \begin{figure*}
  \centering
  \begin{subfigure}[b]{0.32\textwidth}
    \centering
    \includegraphics[width=\textwidth]{{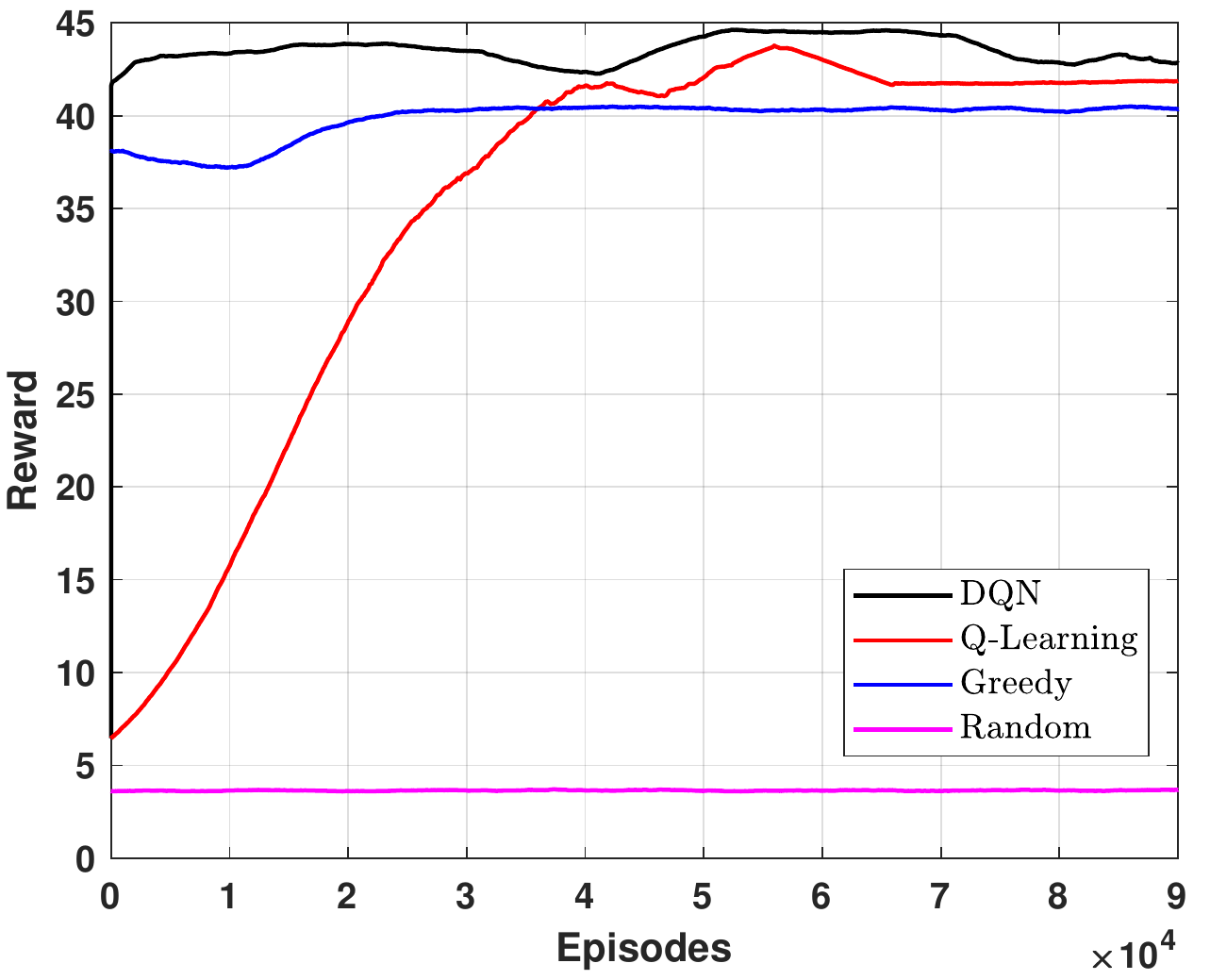}}
    \label{fig:Reward}
  \end{subfigure}
  \begin{subfigure}[b]{0.32\textwidth}
    \centering
    \includegraphics[width=\textwidth]{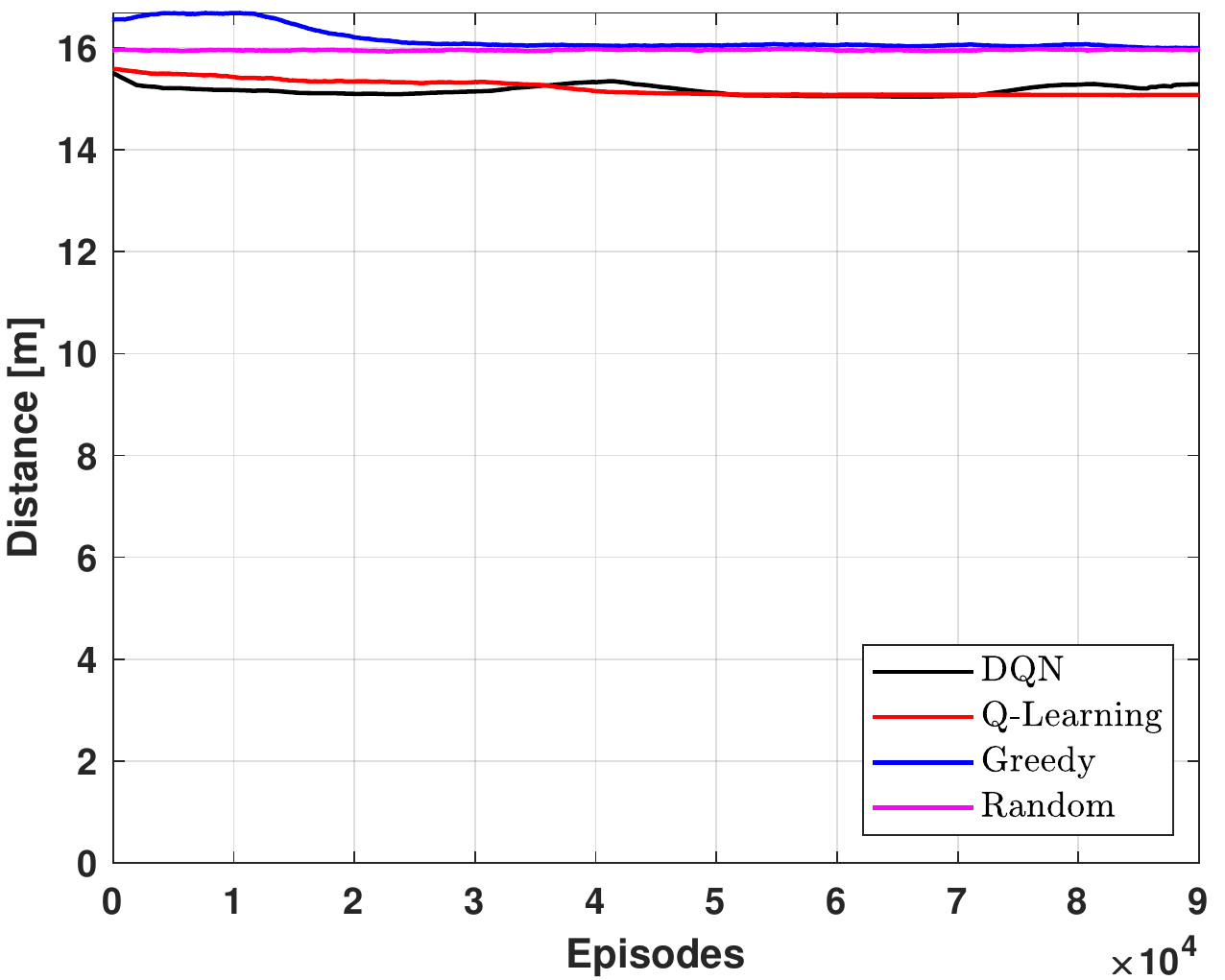}
    \label{fig:Distance}
  \end{subfigure}
  \begin{subfigure}[b]{0.32\textwidth}
    \centering
    \includegraphics[width=\textwidth]{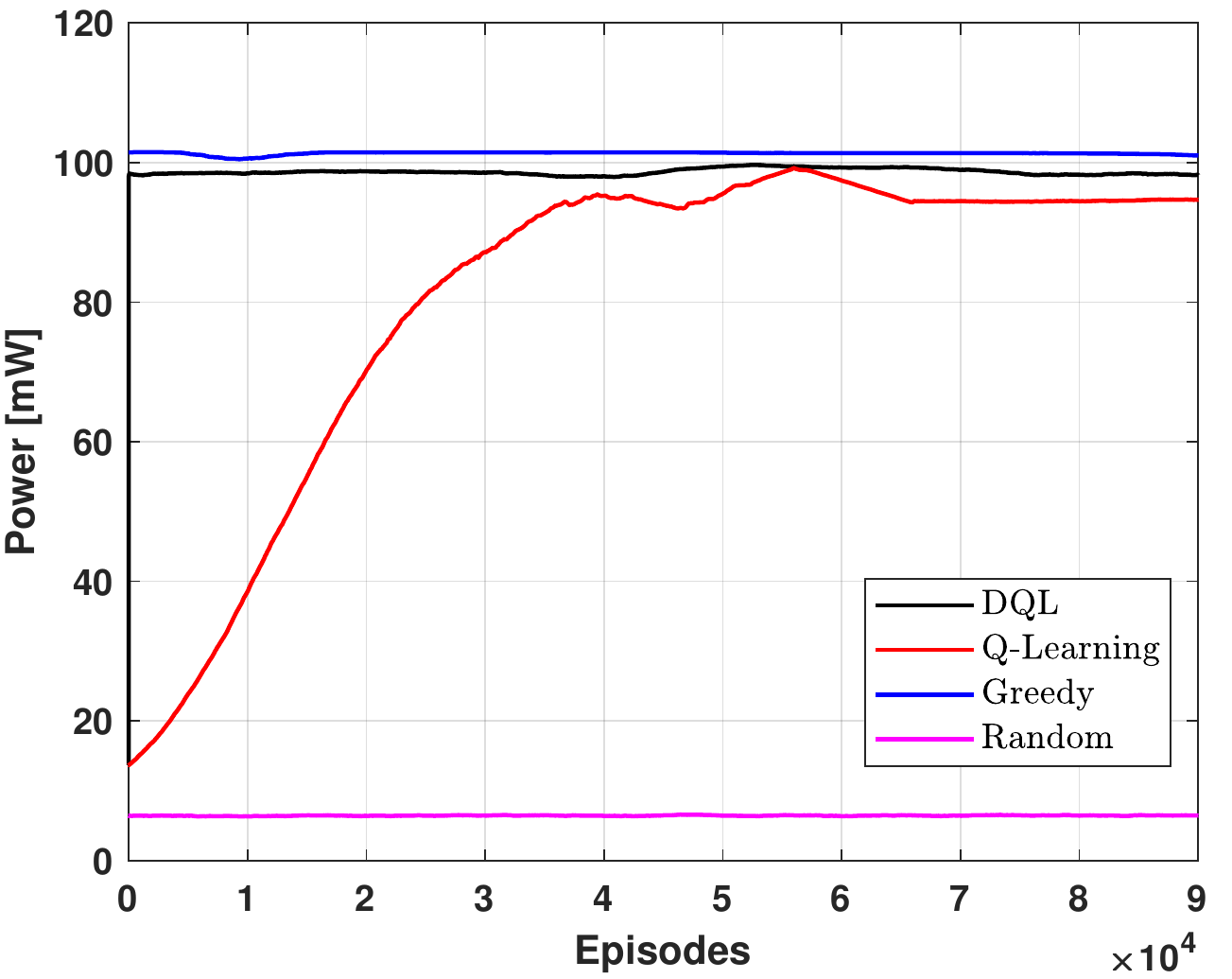}
    \label{fig:PW}
  \end{subfigure}
  \vspace{-5mm}
  \caption{The reward (a), distance (b), and power (c) convergence of the DQN, Q-learning and other base line techniques. }
  \label{fig:Results}
  \vspace{-5mm}
\end{figure*}

Figure~\ref{fig:Results}(a) shows our accumulative reward function over the total number of episodes that is defined as the legitimate UL SNR of the ground UE 
for the proposed DQN and for the other techniques. The illustration of rewards informs that the proposed solution 
has the fastest convergence rate to the highest SNR. 
The SNR performance of the Q-learning is similar, but takes longer to converge because of the nature of the problem that has a large number of states and actions.  

Since the 
optimization problem in~(\ref{eq:opt_secrecy_rate}) relies on the location of the ABS and the transmission power, it is critical to study the convergence of these two parameters. 
Figures~\ref{fig:Results}(b) and~\ref{fig:Results}(c) illustrate the convergence of the UAV position and the transmission power over the number of episodes. 
Comparing the converged position of the ABS and the transmission power for the DQN and Q-learning, we observe that both techniques reach the same ABS position; however, there is a slight increase of the converged transmission power of the DQN over the converged Q-learning transmission power. This difference 
is the main explanation behind the difference in the optimized legitimate SNR levels presented in Fig.~\ref{fig:Results}(a). The low SNR of the random approach 
results from low the transmission power value of Fig.~\ref{fig:Results}(c). Note that the ABS position of the random scheme converges to the same value as the greedy algorithm.

\begin{figure}[t]
     \centering
     \includegraphics[width=0.35\textwidth]{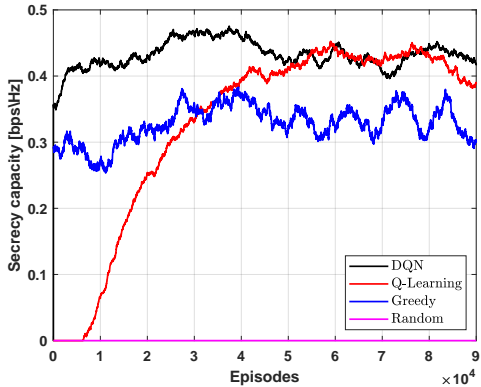}
     \caption{Secrecy capacity achieved by the DQN, Q-learning and baseline techniques.}
     \vspace{-2mm}
     \label{fig:Secrecy}
 \end{figure}

Figure~\ref{fig:Secrecy} shows the secrecy capacity of the DQN compared to the other 
techniques over the number of episodes. 
The secrecy capacity is calculated using the optimized ABS position and transmission power level after finalizing the learning. The proposed DQN algorithm improves the 
secrecy capacity by as much as 40\%, 10\%, and 5\% when compared to the random, greedy, and the Q-learning solutions, respectively. 
The 
DQN reaches a relatively stable secrecy capacity value already after $2 \times 10^4$ episodes. On the other hand, the convergence of Q-learning occurs after $5 \times 10^4$ episodes. This proves the superiority of the DQN regarding the speed of convergence. 
    
\section{Conclusions}
\vspace{-1 mm}
\label{sec:conclusions}

This paper has presented a novel positioning and UL transmission power control approach for ABSs that serve a ground user. 
We consider the legitimate user capacity as the metric to optimize in eavesdropping scenarios without knowledge of the eavesdropper location. This 
contributes to improving the secrecy capacity of the ground user under attack. 
We have provided detailed information about the designed DQN and the Q-learning algorithms.
The obtained results show that the highest capacity and secrecy capacity are achieved with the proposed DQN when compared with Q-learning and two baseline techniques. 

In future research, we will extend the simulation environment to cover multiple legitimate and malicious nodes, \textcolor{black}{analyze the performance of using a multi-antenna systems, and consider both the uplink and the downlink.
} Moreover, the presented technique can be implemented into a testbed, such as 
AERPAW \cite{aerpaw}, which provides robust drones with modular software radio hardware and software, collocated computers, and an experimental license for RF radiation, enabling A2G 
wireless experiments. An AERPAW experiment can deploy ground users and UAVs, where the UAV can implement an ABS that uses the proposed method to position itself for serving a legitimate user and control the transmission power in such a way to maximize the user rate in the presence of an eavesdropper.


\section*{Acknowledgement}
The work of A.~S.~Abdalla 
and
V.~Marojevic  
was supported in part by the NSF 
award CNS-1939334. 

\balance

\bibliographystyle{IEEEtran}
\bibliography{Refs}

\vspace{0.2cm}
\noindent

\vspace{0.2cm}
\noindent

\vspace{0.2cm}
\noindent

\end{document}